\documentclass[showpacs,showkeys,11pt,
preprint,preprintnumbers,nofootinbib,
groupedaddress,superscriptaddress,amsmath,amssymb]{revtex4}

\usepackage[dvipdfm]{graphicx}
\usepackage{cancel,graphics,subfigure}
\usepackage{color}

\newcommand{\SU}{{\text{SU}}}
\newcommand{\GeV}{{\text{GeV}}}
\newcommand{\TeV}{{\text{TeV}}}
\newcommand{\mH}{m_{H^{\pm\pm}}}

\begin{document}
\title{
Discrimination of models including doubly charged scalar bosons\\
by using tau lepton decay distributions
}
\preprint{UT-HET 070}
\author{Hiroaki Sugiyama}
\email{sugiyama@sci.u-toyama.ac.jp}
\affiliation{Department of Physics, University of Toyama, Toyama
930-8555, Japan} 
\author{Koji Tsumura}
\email{ko2@eken.phys.nagoya-u.ac.jp}
\affiliation{Department of Physics, Graduate School of Science, Nagoya
University, Nagoya 464-8602, Japan}
\author{Hiroshi Yokoya}
\email{hyokoya@hep1.phys.ntu.edu.tw}
\affiliation{Department of Physics and Center for Theoretical Sciences, National Taiwan University, Taipei 10617, Taiwan}
\affiliation{National Center for Theoretical Sciences, National Taiwan University, Taipei 10617, Taiwan}



\begin{abstract}
The doubly charged scalar boson ($H^{\pm\pm}$) is introduced in
 several models of the new physics beyond the standard model. 
The $H^{\pm\pm}$ has Yukawa interactions
 with two left-handed charged leptons
 or two right-handed charged leptons
 depending on the models.
We study kinematical properties of $H^{\pm\pm}$ decay products through
 tau leptons in order to discriminate the chiral structures of the new
 Yukawa interaction. 
The chirality of tau leptons can be measured by the energy
 distributions of the tau decay products, and thus the chiral
 structure of the new Yukawa interaction can be traced in
 the invariant-mass distributions of the $H^{\pm\pm}$ decay products. 
We perform simulation studies for the typical decay patterns of the
 $H^{\pm\pm}$ with simple event selections and tau-tagging procedures,
 and show that the chiral structure of the Yukawa interactions of $H^{\pm\pm}$
 can be distinguished by measuring the invariant-mass distributions. 
\end{abstract}
\maketitle

\section{Introduction} 

 The existence of the neutrino masses has been established well%
~\cite{Ref:solar-v,Ref:atom-v,Ref:acc-disapp-v,Ref:acc-app-v,Ref:short-reac-v,Ref:long-reac-v}. 
 However,
neutrinos are massless in the standard model~(SM)
because of the absence of the right-handed partners.
 If the lepton number conservation is violated
in a new physics model beyond the SM,
neutrinos can be Majorana particles
which form a mass term 
with its self-conjugation field only~\cite{Ref:Majorana}, 
since neutrinos are electrically neutral
unlike to all other SM fermions.
 Therefore,
it seems natural to expect
that the possible Majorana nature of neutrinos
provides the reason why neutrinos have very different masses
from those of other SM fermions.

 The doubly charged scalar boson $H^{--}$,
which has a twice electrical charge of the electron,
exists in several models to generate Majorana neutrino masses.
 For instance, 
the particle is a member of an $\SU(2)_L$ triplet scalar field
in the Higgs triplet model~(HTM)~\cite{Ref:Type-II}.
 The triplet field develops a tiny vacuum expectation value~(VEV),
which breaks the lepton number conservation
and is the source of the neutrino mass.
 Such a triplet field appears also
in some models of extended gauge symmetries~\cite{Ref:ExtGauge}.
 On the other hand,
a doubly charged scalar boson is introduced as an $\SU(2)_L$ singlet
scalar field in the Zee-Babu model~(ZBM)~\cite{Ref:Zee-Babu}
which generates Majorana neutrino masses at the two-loop level.
 In these models with $H^{--}$,
its Yukawa interactions with charged leptons
depend on the $\SU(2)_L$ property of $H^{--}$.
 Namely,
$H^{--}$ from an $\SU(2)_L$ triplet field 
couples only with left-handed charged leptons $\ell_L^-$
while the one from an $\SU(2)_L$ singlet field
interacts only with right-handed charged leptons $\ell_R^-$.
Furthermore, $H^{--}$ can be a component of other $\SU(2)_L$
multiplet scalars~\cite{Ref:ExoFermi}, and such $H^{--}$
also has Yukawa interactions with two left-handed
or two right-handed charged leptons
through the mixings between leptons and new fermions. 
 In any case,
both of two charged leptons which couple with $H^{--}$
via the Yukawa interaction
are left-handed or right-handed.
 The discrimination of the chiral structure of the Yukawa interaction
plays an important role to distinguish these models.

 The $H^{\pm\pm}$ can be produced by the pair creation process, 
$pp \to \gamma^\ast/Z^\ast \to H^{++}H^{--}$. 
For $\SU(2)_L$ non-singlet representations, the associated production $pp \to
W^{\mp\ast} \to H^\pm H^{\mp\mp}$ with a singly charged scalar boson
($H^\pm$) is also possible~\cite{Ref:AC-ACG}. 
Theoretical studies for $H^{\pm\pm}$ decaying into same-signed
leptons and weak gauge bosons can be found in, e.g.,
Refs.~\cite{Ref:H++col,Ref:H++WW}. 
The experimental search results for $H^{\pm\pm}$ have been available, 
where purely leptonic decay channels are assumed%
~\cite{Ref:H++TeV,Ref:H++CMS,Ref:H++ATLAS}.
 We comment that 
 these bounds on the $H^{\pm\pm}$ mass are dependent on 
 the production mechanism, the decay branching ratios,
 and the mass spectrum of the scalar boson multiplets~\cite{Ref:Cascade}.
 
 In this letter,
we study the consequence of the chiral structure of the Yukawa interaction
(of the doubly charged scalar boson with two charged leptons) to
the kinematical distribution involving the decay of tau leptons.
The polarization of $\tau$ leptons is known to be probed by its decay
products, and can be exploited to test the structure of new interactions
in the models beyond the SM~\cite{Ref:BHM,Ref:Nojiri,Ref:CHKMZ}.
In Section~II, models of neutrino masses with $H^{\pm\pm}$
are introduced with particular
attention to the chiral structure of the Yukawa interaction. 
In Section~III, the polarization dependences of the decay distributions
of $\tau$ leptons are reviewed, and the invariant-mass distributions 
of final-state particles in the decay of $H^{\pm\pm}$ into at least one 
$\tau$ lepton are discussed. 
Simulation results including $\tau$-tagging and simple kinematical cuts
are also presented. 
Conclusions are given in Section~IV\@.

\section{Models with doubly charged scalar bosons}

 In this section,
we briefly present examples of models
which include the doubly charged scalar boson.

 The first example is the HTM~\cite{Ref:Type-II}.
 In this model,
an $\SU(2)_L$ adjoint scalar field $\Delta$ with hypercharge $Y=1$
is introduced in order to generate masses of neutrinos
via the triplet Yukawa interaction.
 The new Yukawa interaction is given by
\begin{align}
{\mathcal L}_{\text{HTM}}^\text{yukawa}
&= -\overline{L^c} \, h_M \, i\sigma_2 \Delta \, L +\text{H.c.}, 
\end{align}
where $L=(\nu_L^{},\ell_L^-)^T$ is the lepton doublet field, 
the Yukawa coupling matrix is symmetric $h_M^{}=h_M^T$, 
$\sigma_i (i=1\text{-}3)$ are the Pauli matrices, 
and
\begin{align}
\Delta
=
 \begin{pmatrix}
  \Delta^+/\sqrt2
   & \Delta^{++} \\ 
  \Delta^0
   & -\Delta^+/\sqrt2
 \end{pmatrix} .
\end{align}
In the HTM, the doubly charged scalar boson 
interacts with a pair of {\it left-handed} charged leptons. 
 The neutrino mass matrix in the flavor basis is obtained as 
$M_\nu^\text{HTM}
= 2 h_M^\dag \langle \Delta^0 \rangle 
= U_\text{MNS}\, {\widehat M_\nu}\, U_\text{MNS}^T,$
where $\langle\Delta^0\rangle$ is the VEV of the triplet field, 
$\widehat M_\nu$ is the neutrino mass matrix in the diagonal basis, 
and $U_\text{MNS}$ is the Maki-Nakagawa-Sakata~(MNS) matrix
for the lepton flavor mixing. 
 Since the neutrino mass matrix is directly related to the Yukawa matrix, 
the decay patterns of the doubly charged scalar boson 
are constrained by observed neutrino oscillation data~\cite{Ref:HTM}. 
 For example, 
$(h_M^{})_{\mu\mu} \approx (h_M^{})_{\tau\tau}$
and $(h_M^{})_{e\mu} \approx (h_M^{})_{e\tau}$
are required
because the observed neutrino mass matrix approximately 
has the $\mu$-$\tau$ exchange symmetry. 
 By assuming the realistic values of decay branching ratios, constraints
 on the mass of $\Delta^{\pm\pm}$ are 
obtained as $m_{\Delta^{\pm\pm}}^{} \gtrsim 400\,\GeV$~\cite{Ref:H++CMS}.

 The next example is the ZBM~\cite{Ref:Zee-Babu}.
 Two $\SU(2)_L$ singlet scalar bosons,
$k^-$~($Y=-1$) and $k^{--}$~($Y=-2$),
are introduced in the ZBM
to generate tiny neutrino masses at the two-loop level.
 The new interaction terms
which relevant to the radiative neutrino mass
are
\begin{align}
{\mathcal L}_\text{ZBM}
&=
 - \overline{L^c}\, Y_a \,i\sigma_2\, L\, k^+
 - \overline{(\ell_R^-)^c}\, Y_s\, \ell_R^-\, k^{++} 
-\mu\, k^-k^-k^{++} + \text{H.c.}, 
\end{align}
where $Y_a=-Y_a^T$ and $Y_s=Y_s^T$.
 The doubly charged scalar boson in this model
interacts with {\it right-handed} charged leptons. 
 If a lepton number $2$ is assigned to $k^-$ and $k^{--}$,
a coupling constant $\mu$ is the soft breaking parameter
of the lepton number conservation.
 The neutrino mass matrix is calculated as
$\left( M_\nu^\text{ZBM} \right)_{\alpha\beta}
=
 16 \mu\, 
 (Y_a^*)_{\alpha\ell}\, m_{\ell}^{}\,
 (Y_s)_{\ell\ell'}\, I_{\ell\ell'}\,
 m_{\ell'}^{}\, (Y_a^\dag)_{\ell'\beta},$
where the loop function $I_{\ell\ell'}$ is given in Ref.~\cite{Ref:2loopfunc}. 
In order to describe the observed neutrino oscillation parameters, 
$(Y_s^{})_{\mu\mu}(m_\mu/m_\tau)^2 \sim (Y_s^{})_{\mu\tau}(m_\mu/m_\tau) \sim (Y_s^{})_{\tau\tau}$ 
is favored in the ZBM~\cite{Ref:ZBM}.
 This may suggest that
$k^{--} \to \tau_R^-\, \tau_R^-$ would be highly suppressed
while $k^{--} \to \mu_R^-\, \tau_R^-$ could be sizable
in the ZBM\@.
Assuming purely muonic decay mode,
the mass of $k^{\pm\pm}$ is constrained to be 
$m_{k^{\pm\pm}}^{} \gtrsim 250\,\GeV$~\cite{Ref:H++ATLAS}.%
\footnote
{
 If we use theoretical curves in Fig.~2 of Ref.~\cite{Ref:H++ATLAS}
with the result of Ref.~\cite{Ref:H++CMS}
($m_{\Delta^{\pm\pm}} > 391\,\GeV$
for pair-produced $\Delta^{\pm\pm}$
with the 100 \% decay branching ratio into a muon pair.),
we would naively arrive at $m_{k^{\pm\pm}}^{} \gtrsim 320\,\GeV$.
}

\section{Tau polarizations and doubly charged scalar boson decays}

\subsection{Decay distributions of polarized tau leptons}

In this section, we review the polarization dependence of decays of
$\tau$'s, and discuss how that could be traced in the case of
$H^{\pm\pm}$ decay through $\tau$'s. 
In the following discussion, we assume that the leptonic decays of
 doubly charged scalar bosons occur via the Yukawa interactions,
 e.g., 
 $H_X^{--}\overline{\ell_X^{}}(\tau_X^{})^c$ ($X=L, R$), where $H_L^{--}
 (H_R^{--})$ denotes $H^{--}$ only with the left-handed (right-handed)
 interaction.
Hereafter, $\ell$ denotes $e$ or $\mu$. 

First, let us consider the lepton flavor violating~(LFV) decay
$H^{--}\to \ell^- \tau^-$
followed by $\tau^- \to \pi^- \nu$.
The branching ratio of the pionic decay of $\tau$ is about 11\,\%
while the branching ratio of the total hadronic decay is about 65\,\%.
The invariant-mass of $\ell\pi$ is expressed as $M^2_{\ell\pi}=z\, \mH^2$
in the collinear limit, 
where $\mH^{}$ is the mass of $H^{\pm\pm}$ and $z\equiv
E_{\pi}/E_{\tau}$; 
the $E_\pi$ and $E_\tau$ are energies of a pion and a $\tau$ lepton in the
laboratory frame, respectively. 
This relation between the invariant-mass and the energy fraction 
is a good approximation for an energetic $\tau$ lepton, 
e.g., a $\tau$ lepton produced by a heavy particle decay.

The distributions of the pion energy fraction $z$ (namely, of the
invariant-mass $M^2_{\ell\pi}$) are given as
\begin{subequations}\label{Eq:MuPi}
\begin{align}
 & {\cal D}^\pi_{L}(z) = F_{L}^{\pi}(z) = 2(1-z), \label{Eq:MuPi-L} \\
 & {\cal D}^\pi_{R}(z) = F_{R}^{\pi}(z) = 2z, \label{Eq:MuPi-R}
\end{align}
\end{subequations}
where $F_{L,R}^{\pi}(z)$ are the fragmentation functions of
$\tau^-_{L,R}\to\pi^- \nu$ decay in the collinear
limit~\cite{Ref:BHM}.
The fragmentation functions for the other hadronic decay modes
are also known but less sensitive to the polarization of
$\tau$~\cite{Ref:BHM}.
We will utilize these decay modes in the simulation study later.

When the LFV decay is followed by the leptonic decays of $\tau$'s,
the dilepton invariant-mass is expressed as
$M^2_{\ell\ell_\tau} = z\, \mH^2$ in the collinear limit, 
where $z$ is the energy fraction of the
daughter lepton to the parent $\tau$ lepton.
We denote $\ell^{\pm}_\tau$ as $\ell^{\pm}$ from decays
of $\tau^\pm$.%
\footnote{%
The notation $\ell_{(\tau)}\ell_{(\tau)}$ indicates not only
$ee$ and $\mu\mu$ but also $e\mu$.
}
The total branching ratio of the leptonic decays is about 35\,\%.
The distributions of $z$ (namely, of $M^2_{\ell \ell_\tau^{}}$)
are given as
\begin{subequations}\label{Eq:MuEll}
\begin{align}
 & {\cal D}^\ell_{L}(z) = F_{L}^{\ell}(z) = \frac{4}{3}(1-z^3),
 \label{Eq:MuEll-L} \\
 & {\cal D}^\ell_{R}(z) = F_{R}^{\ell}(z) = 2(1-z)^2\left(1+2z\right),
 \label{Eq:MuEll-R}
\end{align}
\end{subequations}
where $F_{L,R}^{\ell}(z)$ are the fragmentation functions of
$\tau_{L,R}^-\to\ell_\tau^- \nu\bar{\nu}$ decay in the collinear
limit~\cite{Ref:BHM}.

\begin{figure}[tb]
 \centering
 \includegraphics[height=5.4cm]{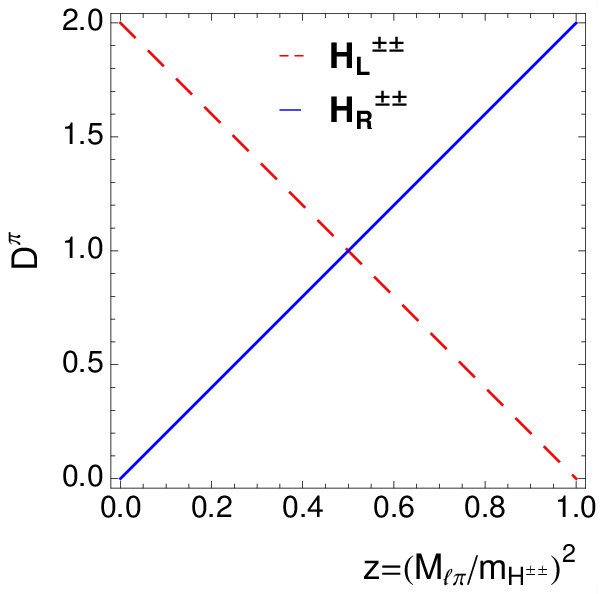}
 \includegraphics[height=5.4cm]{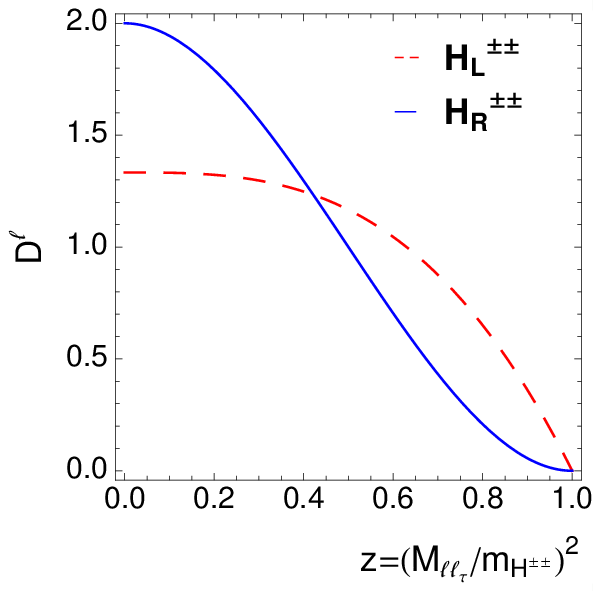}
 \caption{Distributions of the invariant-mass of $\ell\pi$ (left
 panel) and of $\ell\ell_\tau$ (right panel) from the LFV decay 
 $H^{--}\to \ell^- \tau^-$, where $\ell_{(\tau)} = e$,
 $\mu$. 
 The invariant-mass distributions through the decay of $\tau_L$
 ($\tau_R$) are plotted in the dashed (solid) curves.}
 \label{FIG:InvMassMuTau}
\end{figure}

In Fig.~\ref{FIG:InvMassMuTau}, we plot the distributions of the
invariant-masses of $\ell\pi$~(left panel)
and $\ell\ell_\tau$~(right panel)
for the LFV decay $H^{--}\to\ell^- \tau^-$ followed
by the pionic and leptonic decays of the $\tau$, respectively.
The invariant-mass distributions via the decay of $\tau_L$
($\tau_R$) are plotted in the dashed (solid) curves.
For the pionic decay channel, the distributions are linear in $z$ and
have an opposite behavior between $\tau_L$ and $\tau_R$.
On the other hand, the right panel of Fig.~\ref{FIG:InvMassMuTau} shows
that the distribution of $M^{}_{\ell\ell_\tau}$ for the leptonic
$\tau$ decay would be less sensitive to the $\tau$ polarization 
than that of the pionic channel.


Next, we consider the decay mode $H^{--}\to \tau^- \tau^-$.
The decay pattern of the two $\tau$'s can be classified into three
categories: hadronic channels (e.g.\ $\pi\pi$), semi-leptonic channels
(e.g.\ $\ell_\tau\pi$), and purely leptonic channels
($\ell_\tau\ell_\tau$).
The distributions of $\pi\pi$ invariant-mass $M_{\pi\pi}^{}$
in $H^{--}\to\tau^{-}\tau^{-}\to\pi^{-}\pi^{-}\nu\nu$ decay
chain are calculated by convoluting the fragmentation functions of the
pionic decays of $\tau$'s in Eqs.~(\ref{Eq:MuPi}) as follows~\cite{Ref:CHKMZ}: 
\begin{subequations} \label{Eq:PiPi}
\begin{align}
 & {\cal D}^{\pi\pi}_{LL}(z) = \int_{z}^{1} \frac{dz_1}{z_1}\,
 F^{\pi}_{L}(z_1)\, F^{\pi}_{L}\left(z/z_1\right) =
 4\left[(1+z)\log\frac{1}{z}+2z-2\right], \\
 & {\cal D}^{\pi\pi}_{RR}(z) = \int_{z}^{1} \frac{dz_1}{z_1}\,
 F^{\pi}_{R}(z_1)\, F^{\pi}_{R}\left(z/z_1\right) = 4z\log\frac{1}{z},
\end{align}
\end{subequations}
where $z = M^2_{\pi\pi}/\mH^2$ in the collinear limit 
and ${\cal D}^{\pi\pi}_{LL}$~(${\cal
D}^{\pi\pi}_{RR}$) is the distribution for
$H^{\pm\pm}_L$~($H^{\pm\pm}_R$).

The distributions of the $\ell_\tau\pi$ invariant-mass $M_{\ell_\tau\pi}^{}$ for
the $H^{--} \to \tau^- \tau^- \to \ell_\tau^- \pi^- \nu\nu\bar{\nu}$
decay chain are given by 
\begin{subequations}
\begin{align}
 & {\cal D}^{\ell\pi}_{LL}(z) = \int_{z}^{1} \frac{dz_1}{z_1}\,
 F^{\ell}_{L}(z_1)\, F^{\pi}_{L}\left(z/z_1\right) = \frac{4}{9}\left[
 6\log\frac{1}{z}-z^3+9z-8\right], \label{Eq:PiEll-LL}\\
 & {\cal D}^{\ell\pi}_{RR}(z) = \int_{z}^{1} \frac{dz_1}{z_1}\,
 F^{\ell}_{R}(z_1)\, F^{\pi}_{R}\left(z/z_1\right) = 4(1-z)^3,
\end{align}
\end{subequations}
where $z = M^2_{\ell_\tau\pi}/\mH^2$ in the collinear limit. 
The distribution ${\cal D}^{\ell\pi}_{LL}$~(${\cal D}^{\ell\pi}_{RR}$)
is for $H^{\pm\pm}_L$~($H^{\pm\pm}_R$).

The dilepton invariant-mass distributions for the
$H^{--}\to\tau^-\tau^-\to\ell_\tau^-\ell_\tau^-\nu\nu\bar{\nu}\bar{\nu}$
decay chain are given by 
\begin{subequations}
\begin{align}
& {\cal D}^{\ell\ell}_{LL}(z) = \int_{z}^{1} \frac{dz_1}{z_1}\,
 F^{\ell}_{L}(z_1)\, F^{\ell}_{L}\left(z/z_1\right)
= -\frac{16}{27}
 \left[ 2 - 2 z^3 - 3 ( 1 + z^3 ) \log\frac{1}{\,z\,} \right],\\
& {\cal D}^{\ell\ell}_{RR}(z) = \int_{z}^{1} \frac{dz_1}{z_1}\,
 F^{\ell}_{R}(z_1)\, F^{\ell}_{R}\left(z/z_1\right)
 = \frac{4}{\,3\,} \left[
 -5 - 27 z^2 + 32 z^3 + 3 ( 1 + 9 z^2 + 4 z^3 ) \log\frac{1}{\,z\,}
 \right],
\end{align}
\end{subequations}
where $z = M^2_{\ell_\tau\ell_\tau}/\mH^2$ in the collinear limit 
and ${\cal D}^{\ell\ell}_{LL}$~(${\cal
D}^{\ell\ell}_{RR}$) is for $H^{\pm\pm}_L$~($H^{\pm\pm}_R$).

\begin{figure}[tb]
 \centering
 \includegraphics[height=5.4cm]{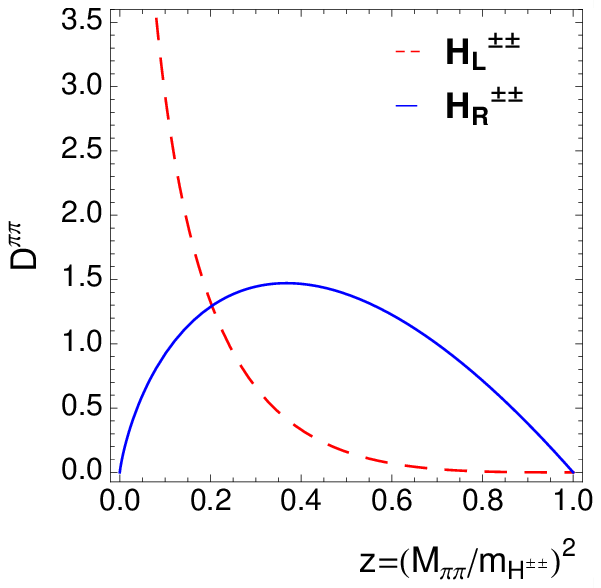}
 \includegraphics[height=5.25cm]{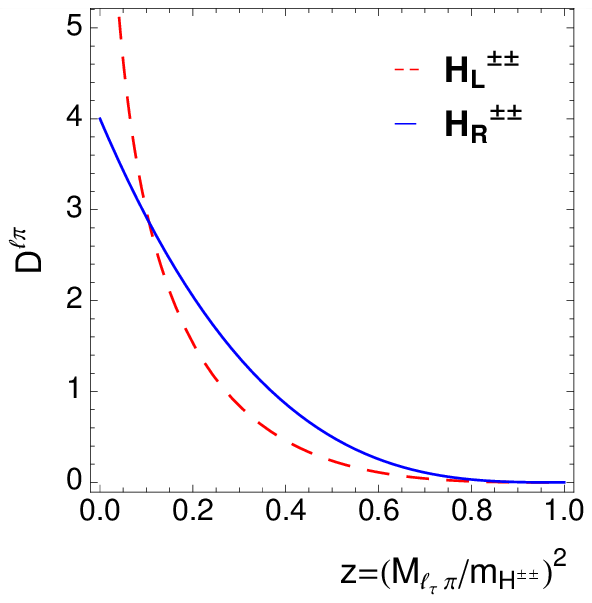}
 \includegraphics[height=5.25cm]{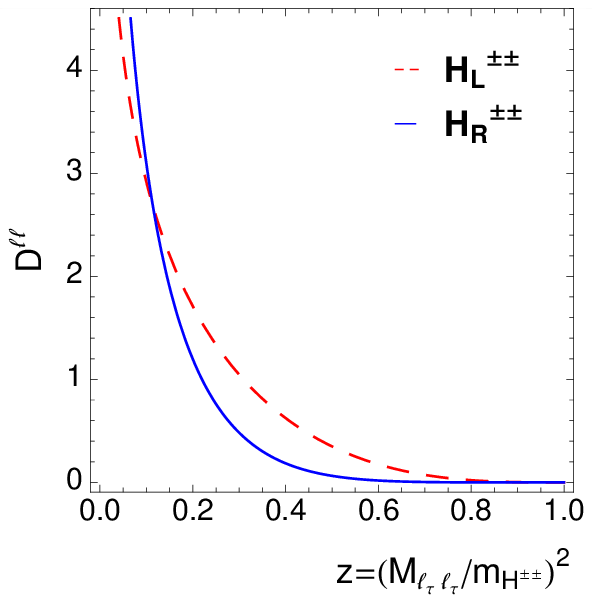}
 \caption{
 Distributions of the invariant-mass
 of $\pi\pi$ (left panel),  $\ell_\tau\pi$ (middle panel) 
 and of $\ell_\tau\ell_\tau$ (right panel)
 from $H^{--}\to\tau^-\tau^-$,
 where $\ell_\tau = e$, $\mu$.
 The invariant-mass distributions through the decay of
 $\tau_L$~($\tau_R$) are plotted
 in the dashed (solid) curves.
 }
 \label{FIG:InvMassTauTau}
\end{figure}

In Fig.~\ref{FIG:InvMassTauTau}, we plot the distributions of the
invariant-mass of $\pi\pi$~(left panel), $\ell_\tau\pi$~(middle
panel), and $\ell_\tau\ell_\tau$~(right panel) from
$H^{--}\to\tau^- \tau^-$.
The invariant-mass distributions through the decay of
$\tau_L$~($\tau_R$) are plotted
in the dashed~(solid) curves. 
Owing to the large polarization dependence of the pionic decay
fragmentation function in Eqs.~(\ref{Eq:MuPi}), the $\pi\pi$
invariant-mass distribution has a large power for the spin analysis.
Namely, the distribution takes a large value
in the region of a small $z$ for the $H_L^{\pm\pm}$ decay while it is large for $z \simeq
0.4$ in the $H_R^{\pm\pm}$ decay.
The $\ell_\tau \pi$ and $\ell_\tau\ell_\tau$ distributions in the
middle and right panels of Fig.~\ref{FIG:InvMassTauTau}, respectively,
take a large value for a small $z$ for both $H_{R}^{\pm\pm}$ and
$H_L^{\pm\pm}$, thus are not useful as the $\tau$ polarization
discriminator.

To summarize, the invariant-mass distributions of $\ell\pi$
(and $\ell\ell_\tau$) in the $H^{--}\to\ell^{-}\tau^{-}$ decay and 
$\pi\pi$ in the $H^{--}\to\tau^{-}\tau^{-}$ decay are good for 
discriminating the $\tau$ polarization and thus the chiral structure
of the Yukawa interaction with $H^{\pm\pm}$.
Notice that the $\ell\pi$~($\ell\ell$) final-state can be reached
through the other decay chain(s) such like
$H^{--}\to\tau^-\tau^-\to\ell_\tau^-\pi^-\nu\nu\bar{\nu}$%
~($H^{--}\to\ell^-\ell^-$ and
$H^{--}\to\tau^-\tau^-\to\ell^-_\tau\ell^-_\tau \nu\nu\bar{\nu}\bar{\nu}$).
If these decay modes for $H^{\pm\pm}$ exist simultaneously,
the signatures from these decay chains would mix in general.
It should be possible to treat the mixed signatures or to divide them by
kinematical cuts.
However, since such analyses depend on the detail branching ratio of
the $H^{\pm\pm}$ decay, we will not consider them in this study.

\subsection{Simulation results}

In order to perform realistic studies for collider experiments, we
examine a Monte-Carlo simulation for the $H^{++}H^{--}$ pair production
and their decays up to parton-showering and hadronizations.
We generate signal events of $pp\to H^{++}H^{--}$ by using {\tt
Pythia}~\cite{Ref:PYTHIA} with handling the $\tau$ decay by {\tt
TAUOLA}~\cite{Ref:TAUOLA} incorporating the chiral properties of the
Yukawa interactions of $H^{\pm\pm}$. 
We consider only the leptonic decay of $H^{\pm\pm}$, and pick up several
patterns of the pair of decays suited for the $\tau$ polarization
measurement. 
The $\mH^{}$ is set to be $400\,\GeV$, and the collider energy to
$\sqrt{s}=14\,\TeV$.
For the reference,
the $H^{++}H^{--}$ production cross-sections is 4.6~fb for the HTM
and 1.9~fb for the ZBM\@. 
For the analysis, we use lighter leptons ($e$ and $\mu$) with
$p_T^{\ell}>15\,\GeV$ and $|\eta_{\ell}|<2.5$,
where $p_T^{}$ is the transverse momentum 
and $\eta$ is the pseudo rapidity. 
To find the hadronically decaying $\tau$'s, we perform $\tau$-tagging 
for every jets with $p^{j}_T>25\,\GeV$ and $|\eta_{j}|<2.5$ which are
constructed by the anti-$k_T$ algorithm~\cite{Ref:Anti-kT} with
$R=0.4$. 
For the $\tau$-tagging, we use two methods.
The first method is devoted to extract the $\tau\to\pi\nu$ decay.
Namely, it is tagged as the pionic $\tau$-jet ($\pi_\tau$) if a jet has
only 1 charged hadron and its transverse energy dominates more than
0.95 of the jet.
The second method is a more general-purpose; we define $j_\tau$ as a jet
which contains 1 or 3 charged tracks in a small cone ($R=0.15$) centered
at the jet momentum direction with the transverse energy deposit to this
small cone more than 0.95 of the jet.
The second method could tag the $\tau$ decay into $2\pi$ and $3\pi$ 
originated from $\tau \to \rho \nu$ and $a_1\nu$ decays, in addition to
the single $\pi$. 
Thus the tagging efficiency is better than the first method, 
but the spin analysis power is weakened.

The extensive signal-to-background studies including $\tau$'s can
be found, for example, in Ref.~\cite{Ref:H++col,Ref:H++WW}.
Following their results, a clear signal extraction is expected by the
requirement of the same-signed dilepton and possibly a peak in their
invariant-mass. 
Thus, we present the simulation results only for the signal events, but
not for the background events.
Expected background processes are diboson production, $t\bar{t}$ plus one
boson production, etc.

As the first case,
we deal with $pp\to H^{++}H^{--}\to\ell^{+}\ell^{+}\ell^{-}\tau^{-}$.
This decay pattern can be easily identified by requiring the same-signed
 dilepton with a sharp peak in their invariant-mass distribution. 
The mass of $H^{\pm\pm}$ can be clearly obtained from the peak.
 Then, the invariant-mass distribution of the remaining $\ell^{-}$ and
 decay products of $\tau^{-}$ can be used for the
 polarization discriminant.
 We take the hadronic decays of $\tau$, 
because of the better spin analysis power than
 the leptonic ones as shown in Fig.~\ref{FIG:InvMassMuTau}.

\begin{figure}[tb]
 \centering
 \includegraphics[height=5.4cm]{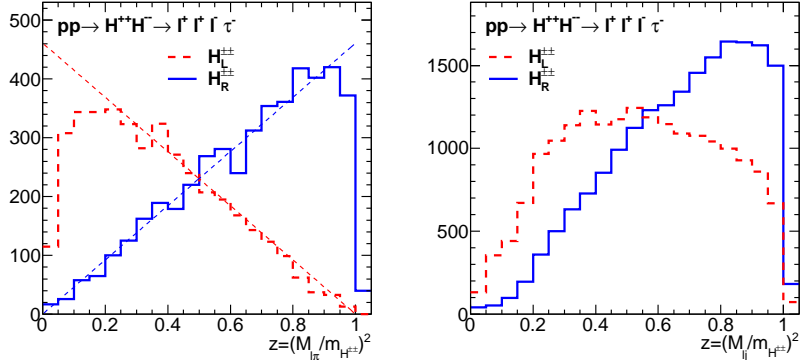}
 \caption{Invariant-mass distributions of $\ell\pi_\tau$ (left) and
 $\ell j_\tau$ (right) in the $pp\to
 H^{++}H^{--}\to\ell^{+}\ell^{+}\ell^{-}\tau^{-}$ process followed by
 hadronic decays of $\tau^{-}$, after the requirement of the same-signed
 dilepton with $M_{\ell\ell}^{}\simeq \mH$.
 Dashed (Solid) histograms are for $H_L^{\pm\pm}$ ($H_R^{\pm\pm}$).
 Smooth lines in the left panel are theoretical expectations by using
 Eqs.~(\ref{Eq:MuPi}) with some normalization.} 
 \label{FIG:SimMuPi}
\end{figure}

In Fig.~\ref{FIG:SimMuPi}, the invariant-mass distributions of
$\ell\pi_\tau$~(left panel) and those of $\ell j_\tau$~(right panel)
from $H^{--}\to \ell^-\tau^-$ are shown for the events with
$3\ell$ and one $\tau$-tagged jet.
 Results for the decay of $H_L^{\pm\pm}$
($H_R^{\pm\pm}$) are plotted in the dashed (solid) histograms. 
The $y$-axis of the plots hereafter stands for the number of events in
our simulation. 
 For each of $H^{\pm\pm}_L$ and $H^{\pm\pm}_R$,
we generate $3\times10^4$ events
of the $pp \to H^{++} H^{--} \to \ell^+ \ell^+ \ell^- \tau^-$ process
followed only by hadronic $\tau$ decays.
 Corresponding integrated luminosity of our simulation
depends on the branching ratio of $H^{\pm\pm}$,
which however we don't specify in this study.
 The signal selection efficiencies
with the $\pi_\tau$ method and the $j_\tau$ method
in Fig.~\ref{FIG:SimMuPi}
are about $13\%$~($15\%$) and $62\%$ ($62\%$),
for $H^{\pm\pm}_L$~($H^{\pm\pm}_R$) events in our simulation,
respectively.

The distributions in the left panel roughly reproduce
the linear dependence on $z$ given in Eqs.~(\ref{Eq:MuPi})
which are superimposed
with some normalization. 
Thus, the pionic $\tau$-tagging method seems to be working well to catch
the pionic $\tau$ decay products. 
The effect of the kinematical cuts can appear in the small $z$ region
due to the $p_T^{}$ cut. 
The discrimination of the two distributions may be possible with just 
a small number of events.
When we employ the general $\tau$-tagging method, the expected number of
events becomes 5 times larger than that for the case with pionic
$\tau$-tagging method. 
Since the distributions for $H_L^{\pm\pm}$ and $H_R^{\pm\pm}$ still
differ from each other, although the difference is weakened, this
general $\tau$-tagging method works as well for our purpose.
We note that the distributions in the right panel could be understood as
the sum of the various hadronic $\tau$ decay contributions with proper
polarization dependence~\cite{Ref:BHM}.
The distributions are scaled by the mass of $H^{\pm\pm}$, thus it is
expected that these distributions does not depend on the value of the
mass so much.
We have confirmed that quite similar distributions are obtained for
the case with a heavier mass of $H^{\pm\pm}$.
\begin{figure}[tb]
 \centering
\includegraphics[height=5.4cm]{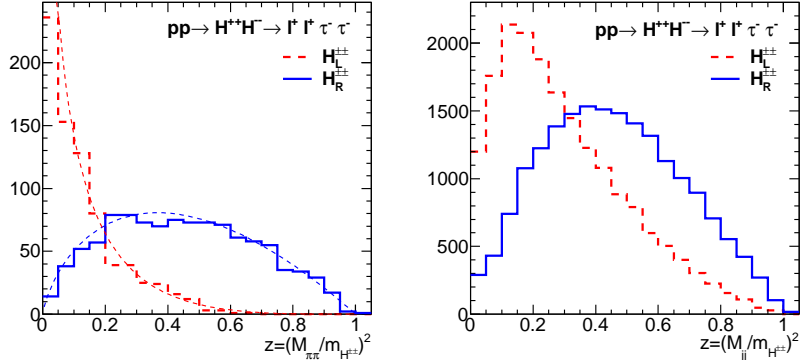}
 \caption{Invariant-mass distributions of $\pi_\tau\pi_\tau$ (left) and
 $j_\tau j_\tau$ (right) in the $pp\to
 H^{++}H^{--}\to\ell^{+}\ell^{+}\tau^{-}\tau^{-}$ process followed by 
 hadronic decays of $\tau^{-}$'s, after the requirement of the
 same-signed dilepton with $M_{\ell\ell}^{}\simeq \mH$.
 Dashed (Solid) histograms are for $H_L^{\pm\pm}$ ($H_R^{\pm\pm}$).
 Smooth lines in the left panel are theoretical expectations by using
 Eqs.~(\ref{Eq:PiPi}) with some normalization.} 
 \label{FIG:SimPiPi}
\end{figure}

 The second case
is $pp\to H^{++}H^{--}\to\ell^{+}\ell^{+}\tau^{-}\tau^{-}$.
We note that
this process can be useful 
when the branching ratio for the LFV decay $H^{--}\to\ell^-\tau^-$ is 
so small that we cannot use the first case ($H^{++}H^{--}\to\ell^{+}\ell^{+}\ell^{-}\tau^{-}$).
This collider signature has also a sharp peak in the invariant-mass
 distribution of the same-signed dilepton, thus the signal event
 extraction from the background contributions would be easy.
We analyze the case where both of $\tau$ leptons decay into hadrons
because of the best spin analysis power as shown in Fig.~\ref{FIG:InvMassTauTau}.

In Fig.~\ref{FIG:SimPiPi}, the invariant-mass distributions of
$\pi_\tau\pi_\tau$ (left panel) and those of $j_\tau j_\tau$ (right
panel) from $H^{--}\to \tau^-\tau^-$ in the $pp\to
H^{++}H^{--}\to\ell^{+}\ell^{+}\tau^{-}\tau^{-}$ process are shown. 
 We generate $3\times 10^{4}$ events of the process
followed only by the hadronic decays of $\tau$'s. 
Results for the decay of $H_L^{\pm\pm}$
($H_R^{\pm\pm}$) are plotted in the dashed (solid) histograms. 
 Distributions of Eqs.~(\ref{Eq:PiPi}) are superimposed to the left panel
with some normalization as references. 
 As expected,
behaviors of the distributions in the left panel are almost the same as
Eqs.~(\ref{Eq:PiPi})
even after the selection cuts.
For the $j_\tau j_\tau$ invariant-mass distributions, the curves for
$H^{\pm\pm}_L$ and for $H^{\pm\pm}_R$ are still well separated.

 The third case is for a situation
in which $H^{--} \to \ell^-\ell^-$ does not exist.
Then the signature would be $pp\to
H^{++}H^{--}\to\ell^{+}\tau^{+}\ell^{-}\tau^{-}$, 
which seems better than
$pp\to H^{++}H^{--}\to\ell^{+}\tau^{+}\tau^{-}\tau^{-}$.
The momentum reconstruction of the two $\tau$'s
is still possible by using 
a collinear approximation for the decay of the two
$\tau$'s~\cite{Ref:H++col}.
 Then the mass of $H^{\pm\pm}$ can be measured
by the invariant-mass of the same-signed $\ell\tau$ pairs.
 If both of the $\tau$'s decay hadronically,
we may suffer from the background contribution.
 In order to suppress the background contribution,
we require that one $\tau$ decays leptonically
which gives the same-signed dilepton $\ell\ell_\tau$.
 The hadronic decay for the other $\tau$
will be preferred for the discrimination of the polarization, 
although the $\ell^{+}\ell^{+}_\tau\ell^{-}\ell^{-}_\tau$ signature
may be also exploited.

\begin{figure}[tb]
 \centering
 \includegraphics[height=5.2cm]{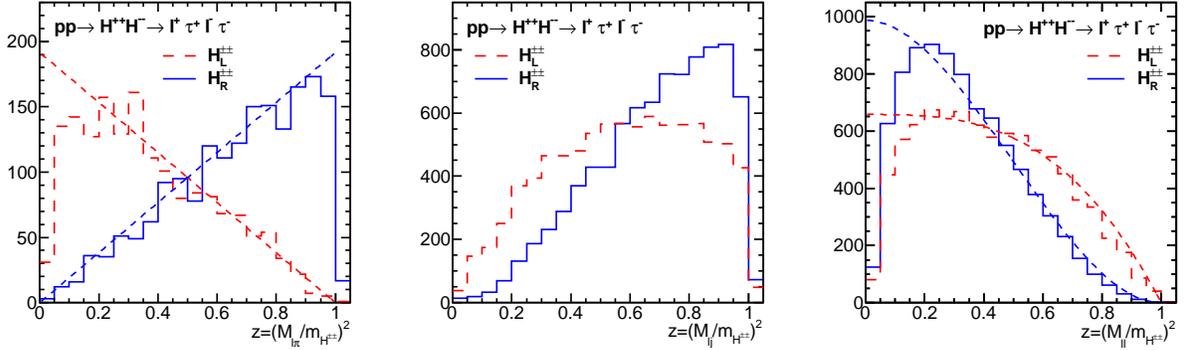}
 \caption{Invariant-mass distributions of $\ell\pi_\tau$ (left), 
 $\ell j_\tau$ (middle) and $\ell\ell_\tau$ (right) in the $pp\to
 H^{++}H^{--}\to\ell^{+}\tau^{+}\ell^{-}\tau^{-}$ process followed by
 one leptonic and one hadronic decays of $\tau$'s after the requirement
 of the proper momentum reconstruction by using the collinear
 approximation method.
 Dashed (Solid) histograms are for $H_L^{\pm\pm}$ ($H_R^{\pm\pm}$).
 Smooth lines in the left and right panels are theoretical expectations
 by using Eqs.~(\ref{Eq:MuPi}) and Eqs.~(\ref{Eq:MuEll}),
 respectively, with some normalization.} 
 \label{FIG:LFV}
\end{figure}

In Fig.~\ref{FIG:LFV}, the invariant-mass distributions of $\ell\pi_\tau$
(left panel), those of $\ell j_\tau$ (middle panel) are shown for the
event in which the momentum reconstruction is resolved by using the
collinear approximation method.%
\footnote{%
One may wonder how to select $\ell_\tau^-$
for the collinear approximation method
under the existence of $\ell^-$.
One can find the correct one which gives the smaller difference of the
reconstructed invariant-masses of the two $\ell^{\pm}\tau^{\pm}$ pairs.
}
 We generate $3\times 10^4$ events of
the $pp\to H^{++}H^{--}\to\ell^{+}\tau^{+}\ell^{-}\tau^{-}$ process.
 The decay of a tau lepton is not restricted to the hadronic ones
in contrast with simulations for Figs.~\ref{FIG:SimMuPi} and \ref{FIG:SimPiPi}. 
The full momentum reconstruction is not necessary to obtain the plots in
Fig.~\ref{FIG:LFV}, but is very effective to extract the signal events
of this decay pattern.
 Distributions of Eqs.~(\ref{Eq:PiPi})
are superimposed to the left panel
with some normalization as references. 
 Behaviors of 
the obtained invariant-mass
distributions of $\ell\pi_\tau$ and $\ell j_\tau $
are almost the same as those in the $pp\to 
H^{++}H^{--}\to\ell^{+}\ell^{+}\ell^{-}\tau^{-}$ process in
Fig.~\ref{FIG:SimMuPi}.
In addition, in the right panel, we plot the invariant-mass
distributions of the same-signed $\ell\ell_\tau$ pair from the other
side of the $H^{\pm\pm}$ decays.
The behaviors of $\ell\ell_\tau$ invariant-mass distributions
are in good agreement with Eqs.~(\ref{Eq:MuEll})
whose distributions are superimposed
with some normalization as references. 
Due to the good momentum resolution of leptons, the dilepton
invariant-mass distributions could be also useful for the $\tau$
polarization discrimination.

 Finally,
we comment on the case
where $H^{--}$ decays only into $\ell^-\ell^-$ or $\tau^- \tau^-$.
 If $H^{--}$ decays only into $\ell^-\ell^-$, it is impossible to observe
any polarization phenomena from kinematical measurements.
 Then
we should rely on predictions in each model
about the $H^{\pm\pm}$ production cross-section and
the decay branching ratios
in order to distinguish models.
 If $H^{--}$ decays only into $\tau^-\tau^-$, 
 $H^{++}H^{--}$ gives the $4\tau$ signature.
Since, there are too many sources of the missing momentum, 
 the momentum reconstruction for the $\tau$'s is not possible,
we cannot use the invariant-mass peak at $\mH^{}$
for the background reduction.
 For the similar signatures in the context of the two Higgs doublet model,
however,
a sufficient background reduction is expected in the various
channels for the $\tau$ decays~\cite{Ref:KTY}.
Thus, the invariant-mass distributions, for example, in the
$\ell^{\pm}_\tau\ell^{\pm}_\tau j^{}_\tau j^{}_\tau$ channel may be used
for the polarization discriminant.
Notice that the information on $\mH^{}$ could be obtained by 
the endpoint of the invariant-mass distributions~\cite{Ref:KTY}.


\section{Conclusions}

 The doubly charged scalar boson $H^{\pm\pm}$
appears in several new physics models,
especially in the models to generate Majorana neutrino masses.
 The $H^{\pm\pm}$ has characteristic Yukawa interactions
with two {\it left-handed} charged leptons
(e.g., $\Delta^{\pm\pm}$ from an $\SU(2)_L$ triplet field in the HTM)
or two {\it right-handed} charged leptons
(e.g., $k^{\pm\pm}$ from an $\SU(2)_L$ singlet field in the ZBM)
depending on the model.
 We have studied the kinematical consequences of the Yukawa interactions
in order to discriminate these models
through the determination of the chiral structure of the Yukawa interaction.

 At collider experiments,
it is known that
the polarization of the $\tau$ lepton is analyzed
by the energy fraction distributions of its decay products
($\pi^{\pm}$, $\ell^{\pm}$, etc.). 
We have seen that the invariant-mass distributions are good analyzers
of the $\tau$ polarization
especially for the hadronic $\tau$ decays.

 We have performed a simple Monte-Carlo simulation
for decays of $\tau$'s made from the decays of pair-produced $H^{\pm\pm}$.
 If $H^{++}H^{--} \to \ell^{+} \ell^{+} \ell^{-} \tau^{-}$ mode exists
($\ell = e, \mu$),
the invariant-mass distribution of $\ell^- \pi_\tau^-$
gives the best analysis power on $\tau$ polarization.
 The background contribution can be highly reduced
by requiring $\ell^{+} \ell^{+}$
whose invariant-mass is $\simeq\mH^{}$.
 We have shown that
$H^{++}H^{--} \to \ell^{+} \ell^{+} \tau^{-} \tau^{-}$ mode
followed by $\tau \to \pi \nu$ for both $\tau$
is also useful to discriminate the $\tau$ polarization
by using the distribution of the invariant-mass of $\pi_\tau\pi_\tau$.
 Even if there is no $H^{++} \to \ell^+\ell^+$,
we have found that
$\tau$ polarization can be determined
by the distribution of the invariant-mass $M_{\ell\pi_\tau}$
in $H^{++}H^{--} \to \ell^{+} \tau^{+} \ell^{-} \tau^{-}$ mode
with a leptonic and a pionic $\tau$ decays.
 The reduction of the background events
can be achieved by requiring same-signed $\ell\ell_\tau$
with the collinear approximation method.
Therefore,
in these various cases for the leptonic decay of $H^{\pm\pm}$,
we can determine the chiral structure
of the Yukawa interaction of $H^{\pm\pm}$,
and it will help to discriminate new physics models
beyond the SM\@.

\acknowledgments
The work of H.S.\ was supported in part by the Grant-in-Aid for Young
Scientists (B) No.~23740210. 
The work of K.T.\ was supported, in part, by the Grant-in-Aid for
Scientific research from the Ministry of Education, Science, Sports, and
Culture (MEXT), Japan, No.23104011. 
The work of H.Y.\ was supported in part by the National Science Council
of Taiwan under Grant No.~NSC 100-2119-M-002-001. 



\begin{thebibliography}{99}

\bibitem{Ref:solar-v}
  B.~T.~Cleveland {\it et al.},
  Astrophys.\ J.\  {\bf 496}, 505 (1998);
%
  W.~Hampel {\it et al.}  [GALLEX Collaboration],
  Phys.\ Lett.\  B {\bf 447}, 127 (1999);
%
  B.~Aharmim {\it et al.}  [SNO Collaboration],
  Phys.\ Rev.\ Lett.\  {\bf 101}, 111301 (2008);
%
  J.~N.~Abdurashitov {\it et al.}  [SAGE Collaboration],
  Phys.\ Rev.\  C {\bf 80}, 015807 (2009);
%
  K.~Abe {\it et al.}  [Super-Kamiokande Collaboration],
  Phys.\ Rev.\  D {\bf 83}, 052010 (2011);
%
  G.~Bellini {\it et al.} [Borexino Collaboration],
  Phys.\ Rev.\ Lett.\  {\bf 107}, 141302 (2011).

\bibitem{Ref:atom-v}
  R.~Wendell {\it et al.}  [Kamiokande Collaboration],
  Phys.\ Rev.\  D {\bf 81}, 092004 (2010).

\bibitem{Ref:acc-disapp-v}
  M.~H.~Ahn {\it et al.}  [K2K Collaboration],
  Phys.\ Rev.\  D {\bf 74}, 072003 (2006);
%
  P.~Adamson {\it et al.}  [The MINOS Collaboration],
  Phys.\ Rev.\ Lett.\  {\bf 106}, 181801 (2011).

\bibitem{Ref:acc-app-v}
  K.~Abe {\it et al.}  [T2K Collaboration],
  Phys.\ Rev.\ Lett.\  {\bf 107}, 041801 (2011).

\bibitem{Ref:short-reac-v}
  M.~Apollonio {\it et al.}  [CHOOZ Collaboration],
  Eur.\ Phys.\ J.\  C {\bf 27}, 331 (2003);
%
  Y.~Abe {\it et al.}  [DOUBLE-CHOOZ Collaboration],
  Phys.\ Rev.\ Lett.\  {\bf 108}, 131801 (2012);
%
  F.~P.~An {\it et al.}  [DAYA-BAY Collaboration],
  Phys.\ Rev.\ Lett.\  {\bf 108}, 171803 (2012);
%
  J.~K.~Ahn {\it et al.}  [RENO Collaboration],
  Phys.\ Rev.\ Lett.\  {\bf 108}, 191802 (2012).

\bibitem{Ref:long-reac-v}
  A.~Gando {\it et al.}  [The KamLAND Collaboration],
  Phys.\ Rev.\  D {\bf 83}, 052002 (2011).

\bibitem{Ref:Majorana}
  E.~Majorana,
  Nuovo Cim.\  {\bf 14}, 171 (1937).
  
\bibitem{Ref:Type-II}
  W.~Konetschny and W.~Kummer,
  Phys.\ Lett.\  B {\bf 70}, 433 (1977);
%
  M.~Magg and C.~Wetterich,
  Phys.\ Lett.\  B {\bf 94}, 61 (1980);
%
  T.~P.~Cheng and L.~F.~Li,
  Phys.\ Rev.\  D {\bf 22}, 2860 (1980);
%
  J.~Schechter and J.~W.~F.~Valle,
  Phys.\ Rev.\  D {\bf 22}, 2227 (1980).

\bibitem{Ref:ExtGauge} 
  R.~N.~Mohapatra and G.~Senjanovic,
  Phys.\ Rev.\ Lett.\  {\bf 44}, 912 (1980);
%
  G.~Lazarides, Q.~Shafi and C.~Wetterich,
  Nucl.\ Phys.\ B {\bf 181}, 287 (1981);
%
  N.~Arkani-Hamed, A.~G.~Cohen, E.~Katz and A.~E.~Nelson,
  JHEP {\bf 0207}, 034 (2002).

\bibitem{Ref:Zee-Babu}
  A.~Zee,
  Nucl.\ Phys.\  B {\bf 264}, 99 (1986).
%
  K.~S.~Babu,
  Phys.\ Lett.\  B {\bf 203}, 132 (1988).

\bibitem{Ref:ExoFermi}
  M.~Aoki, S.~Kanemura and K.~Yagyu,
  Phys.\ Lett.\ B {\bf 702}, 355 (2011)
  [Erratum-ibid.\ B {\bf 706}, 495 (2012)];
%
  B.~Ren, K.~Tsumura and X.~-G.~He,
  Phys.\ Rev.\ D {\bf 84}, 073004 (2011). 

\bibitem{Ref:AC-ACG}
  A.~G.~Akeroyd and M.~Aoki,
  Phys.\ Rev.\ D {\bf 72}, 035011 (2005);
%
 A.~G.~Akeroyd, C.~-W.~Chiang,
  Phys.\ Rev.\ D {\bf80}, 113010 (2009);
%
  A.~G.~Akeroyd, C.~-W.~Chiang, N.~Gaur,
  JHEP {\bf 1011}, 005 (2010).

\bibitem{Ref:H++col}
  A.~Hektor, M.~Kadastik, M.~Muntel, M.~Raidal and L.~Rebane,
  Nucl.\ Phys.\ B {\bf 787} (2007) 198.

\bibitem{Ref:H++WW}
  P.~Fileviez Perez, T.~Han, G.~-y.~Huang, T.~Li and K.~Wang,
  Phys.\ Rev.\ D {\bf 78}, 015018 (2008); 
%
  C.~-W.~Chiang, T.~Nomura and K.~Tsumura,
  Phys.\ Rev.\ D {\bf 85}, 095023 (2012).

\bibitem{Ref:H++TeV}
D.~Acosta {\it et al.}  [CDF Collaboration],
  Phys.\ Rev.\ Lett.\  {\bf 93}, 221802 (2004);
%
  Phys.\ Rev.\ Lett.\  {\bf 95}, 071801 (2005);
%
T.~Aaltonen {\it et al.}  [CDF Collaboration],
  Phys.\ Rev.\ Lett.\  {\bf 101}, 121801 (2008);
%
  V.~M.~Abazov {\it et al.}  [D0 Collaboration],
  Phys.\ Rev.\ Lett.\  {\bf 93}, 141801 (2004); 
%
  Phys.\ Rev.\ Lett.\  {\bf 101}, 071803 (2008);
%
  Phys.\ Rev.\ Lett.\  {\bf 108}, 021801 (2012).
  
\bibitem{Ref:H++CMS}
CMS Collaboration, CMS PAS HIG-12-005 (March 2011).

\bibitem{Ref:H++ATLAS}
  G.~Aad {\it et al.}  [ATLAS Collaboration],
  Phys.\ Rev.\ D {\bf 85}, 032004 (2012).

\bibitem{Ref:Cascade}
  A.~G.~Akeroyd and H.~Sugiyama,
  Phys.\ Rev.\ D {\bf 84}, 035010 (2011); 
%
  M.~Aoki, S.~Kanemura and K.~Yagyu,
  Phys.\ Rev.\ D {\bf 85}, 055007 (2012);
%
  A.~G.~Akeroyd, S.~Moretti and H.~Sugiyama,
  Phys.\ Rev.\ D {\bf 85}, 055026 (2012).

\bibitem{Ref:BHM}
  B.~K.~Bullock, K.~Hagiwara and A.~D.~Martin,
  Phys.\ Rev.\ Lett.\  {\bf 67}, 3055 (1991);
%
  Phys.\ Lett.\ B {\bf 273}, 501 (1991);
%
  Nucl.\ Phys.\ B {\bf 395}, 499 (1993).

\bibitem{Ref:Nojiri}
  M.~M.~Nojiri,
  Phys.\ Rev.\ D {\bf 51}, 6281 (1995);
%
  M.~M.~Nojiri, K.~Fujii and T.~Tsukamoto,
  Phys.\ Rev.\ D {\bf 54}, 6756 (1996).

\bibitem{Ref:CHKMZ} 
  S.~Y.~Choi, K.~Hagiwara, Y.~G.~Kim, K.~Mawatari and P.~M.~Zerwas,
  Phys.\ Lett.\ B {\bf 648}, 207 (2007).

\bibitem{Ref:HTM}
  E.~J.~Chun, K.~Y.~Lee and S.~C.~Park,
  Phys.\ Lett.\ B {\bf 566}, 142 (2003);
%
  J.~Garayoa and T.~Schwetz,
  JHEP {\bf 0803}, 009 (2008); 
%
  A.~G.~Akeroyd, M.~Aoki and H.~Sugiyama,
  Phys.\ Rev.\ D {\bf 77}, 075010 (2008);
%
  M.~Kadastik, M.~Raidal and L.~Rebane,
  Phys.\ Rev.\ D {\bf 77}, 115023 (2008).

\bibitem{Ref:2loopfunc}
  K.~L.~McDonald and B.~H.~J.~McKellar,
  hep-ph/0309270.

\bibitem{Ref:ZBM}
  M.~Nebot, J.~F.~Oliver, D.~Palao and A.~Santamaria,
  Phys.\ Rev.\ D {\bf 77}, 093013 (2008).
  
\bibitem{Ref:PYTHIA}
  T.~Sjostrand, S.~Mrenna and P.~Z.~Skands,
  JHEP {\bf 0605}, 026 (2006).

\bibitem{Ref:TAUOLA}
  S.~Jadach, Z.~Was, R.~Decker and J.~H.~Kuhn,
  Comput.\ Phys.\ Commun.\  {\bf 76}, 361 (1993).

\bibitem{Ref:Anti-kT}
  M.~Cacciari, G.~P.~Salam and G.~Soyez,
  JHEP {\bf 0804} (2008) 063.

\bibitem{Ref:KTY}
  S.~Kanemura, K.~Tsumura and H.~Yokoya,
  Phys.\ Rev.\ D {\bf 85} (2012) 095001.






\end{thebibliography}
\end{document}